\documentclass[aps,
twocolumn,superscriptaddress]{revtex4-1}

\usepackage[pagewise]{lineno}
\usepackage{color}
\usepackage{amsmath}
\usepackage{graphicx}   
\usepackage{breqn}
\usepackage{natbib}
\usepackage{amssymb}
\usepackage{hyperref}
\newcommand{\comment}[1]{}

\begin{document}
\title{Sub-TeV hadronic interaction model differences and their impact on air showers}

\author{\'A. Pastor-Guti\'errez}
\affiliation{Max-Planck-Institut f\"ur Kernphysik P.O. Box 103980$,$ D 69029$,$ Heidelberg$,$ Germany}

\author{H. Schoorlemmer}
\affiliation{Max-Planck-Institut f\"ur Kernphysik P.O. Box 103980$,$ D 69029$,$ Heidelberg$,$ Germany}

\author{R.D. Parsons} 

\affiliation{Institut f\"ur Physik Humboldt-Universit\"at zu Berlin Newtonstr. 15$,$D 12489$,$ Berlin$,$ Germany}
\affiliation{Max-Planck-Institut f\"ur Kernphysik P.O. Box 103980$,$ D 69029$,$ Heidelberg$,$ Germany}

\author{M.Schmelling}
\affiliation{Max-Planck-Institut f\"ur Kernphysik P.O. Box 103980$,$ D 69029$,$ Heidelberg$,$ Germany}

\date{April 2021}

\begin{abstract}
In the sub-TeV regime, the most widely used hadronic interaction models disagree significantly in their predictions for post-first interaction and ground-level particle spectra from cosmic ray induced air showers. These differences generate an important source of systematic uncertainty in their experimental use.  We investigate the nature and impact of model uncertainties through a simultaneous analysis of ground level particles and first interaction scenarios. We focus on air shower primaries with energies close to the transition between high and low energy hadronic interaction models, where the dissimilarities have been shown to be the largest and well within the range of accelerator measurements. Interaction models are shown to diverge as several shower scenarios are compared, reflecting intrinsic differences in the model theoretical frameworks.   Finally, we discuss the importance of interactions in the energy regime where the switching between models occurs ($<1$\,TeV) and the effect of the choice of model on the number of  hadronic interactions within cosmic ray induced air showers of higher energies.
\end{abstract}
 
\maketitle

\section{Introduction}
High energy particles of cosmic origin interact with our atmosphere and create cascades of secondary particles known as extensive air showers (EASs). These EASs can be observed by recording the fluorescence, Cherenkov, and radio emission produced while propagating through the atmosphere and/or detecting the secondary particles that reach ground level. In order to predict and understand the EAS observations the interactions within the atmosphere must be compared to Monte Carlo simulations commonly performed with software packages like CORSIKA \cite{corsika}. This simulation package bundles electromagnetic and several selectable hadronic interaction packages to calculate the development of the particle cascades. The hadronic interaction event generators are separated in low and high energy models and in the simulation package, one or the other is selected depending on the energy of the interaction being considered. Per default, the energy at which the switch between high and low energy models occurs is 80\,GeV.

In a recent study \cite{Sys_Diff_HIM}, it was found that for cosmic-ray protons with an energy just above the typical switching energy, the properties of the simulated EASs have a strong dependency on the selection of the hadronic interaction model. With increasing energy of the incident proton (up to 100 TeV), the average differences in air shower properties between the models seemed to reduce. In this follow-up study, we focus on the switching energy domain, where the differences between the models are most prominently exposed (up to $60\%$ difference in the ground level observables). We compare in detail the nature of the early shower development and its relation to ground-level observables for different hadronic interaction models.

The differences in ground-level observables at 100\,GeV are driven by the early shower development and can therefore expose intrinsic differences between the hadronic interaction models.  Therefore, low energy showers serve as a laboratory for comparison of differences in the first interactions from the ground level outcome. Another domain where the first interaction will play an important role is in the estimation of background rates for gamma ray astronomy. A fraction of hadronic first interactions will generate an energetic $\pi^0$, which will subsequently generate an electromagnetic cascade, mimicking a gamma-ray induced air shower. The rates of these events generated by different hadronic interaction models has a significant impact on sensitivity studies for gamma-ray observatories \cite{CTA_HADR}. Ideally, the different models should be directly compared to measurements made at dedicated experiments at  particle accelerators (for example \cite{Dembinski_proton_Oxigen}).

\section{Simulation methodology}
All simulations for this study were performed with the Monte Carlo air shower event generator CORSIKA v7.64. The hadronic models tested were EPOS-LHC \cite{EPOS-LHC}, QGSJetII-04 \cite{QGSJetII-04}, SIBYLL 2.3c \cite{Sibyll2.3c,SIBYLL2.3-2020} and UrQMD \cite{UrQMD}.

The transition energy is not an a priori defined parameter and can flexibly be set up to a few hundred GeV. Nevertheless, it was set to 80\,GeV for all high energy models (default value in CORSIKA). UrQMD in contrast is a low energy model and rules the hadronic interactions in air showers below the transition energy. 

The main goal of this research is to understand the hadronic interaction models' deviations in the sub-TeV regime, where they have been seen to significantly disagree \cite{Sys_Diff_HIM}. We chose an initial proton energy of 100\,GeV (including the rest mass) to perform all simulations. 
Since the energy range studied is within the validity region of the transition energy, agreement was expected between using only the low energy model and using the combination of high- and low-energy models. UrQMD's full shower simulations were carried out by setting the transition energy to values larger than the initial proton energy.  

In order to remove the effect of model intrinsic cross-section differences (given in Appendix \ref{Cross-sectionSection}) and track the particles produced in the first interaction, all simulations were performed with a fixed first interaction altitude. We set the altitude of the first interaction  to $17.55$\,km  which corresponds to the average cross-section of the interaction models. The atmospheric model chosen was GDAS/May, corresponding to an atmospheric depth of 85\,g\,cm$^{-2}$ at the initial altitude. Moreover, the collision setup consisted of a fixed target nitrogen nucleus and a zero degree zenith angle incoming proton. The nucleon-nucleon centre-of-mass energy is $\sqrt{s_{NN}} \approx 14$\,GeV.

The most important feature in the simulations was the possibility of observing event-by-event the final state immediately after the first interaction and at ground level. It allowed us to compare ground-level observables from different models for fixed classes of first interactions. Two observational points were defined; the first, 1\,cm below the initial interaction point to register the initial particle production; and the second at the ground level of the HAWC gamma-ray observatory at an altitude of 4100\,m \cite{HAWC_CRAB}. We used the lateral distribution functions (LDF) of the muon and electromagnetic (EM) components as the main observables at ground level.

\section{First Interaction Contributions to the Lateral Distribution Functions}
\label{sec:FirstInteraction}
\subsection{First interaction classification}
To relate the initial collisions with ground level observables, the different first interaction scenarios will be characterised. As a huge diversity of particles can be produced in the hadronic interaction, we group particles into four different ``families" (see Table \ref{particlefamiliestable}). The classification is motivated by the family's impact on the shower development and the preservation of a causal correlation between first interaction products and ground level observables, so for example the \emph{muonic family} contains muons and particles which typically decay to produce muons. The \emph{other hadrons} family was defined to contain particles which have decay channels that simultaneously contribute to multiple components. It is important to separate these hadrons in order to avoid mistaken links between the post-first interaction particle spectrum and ground level observable particle type.

\begin{table}[hbt!] 
\centering
\caption{Definition of particle families.}
\begin{tabular}{|l|l|}
\hline
\textbf{Nucleons  } &  $p$, $\bar{p}$, $n$, $\bar{n}$    \\ \hline
\textbf{Muonic family  } & $\mu^{+}$, $\mu^{-}$, $\pi^{+}$, $\pi^{-}$,  $K^{0}_{L}$, $K^{+}$, $K^{-}$, $K^{0}_{S}$ \\ \hline
\textbf{EM component  } & $\gamma$, $e^{-}$, $e^{+}$\\ \hline
\textbf{Other hadrons  } & $\Lambda$, $\Sigma^{+}$, $\Sigma^{-}$, $\overline{\Sigma}^{-}$, $\overline{\Sigma}^{+}$, $\Xi^{0}$, $\Xi^{-}$, $\Omega^{-}$, $\bar{\Lambda}$, ... \\ \hline
\end{tabular}
\label{particlefamiliestable}
\end{table}

A common parameter used to characterise the energy distributions in an interaction is the inelasticity,
\begin{equation}
    \kappa=1-\frac{E_{\text{LP}}}{E_{\text{FI}}} \approx 1 - x^{\text{LP}}_F,
\end{equation}
where $E_{\text{FI}}$ is the total shower energy (sum of all particle energies) after the first interaction and $E_{\text{LP}}$ the leading particle energy. The inelasticity $\kappa$ is closely related to Feynman's scaling variable  $x^{\text{LP}}_F$ and is a measure for how much energy is available for the production of secondary particles.  The leading particle jointly with the variable $\kappa$ allows  one to identify similar types of interactions within the different models.

Note that the total shower energy after the first interaction is used in the definition of the inelasticity to ensure that $\kappa$ is in the range $[0,1]$. Normalisation to the energy of the initial system (proton plus target) would violate this constraint, since in all four models some violation of energy-momentum  conservation was observed.
This was additionally checked at generator level using the CRMC  software package \cite{CRMC} for EPOS-LHC, QGSJetII-04 and SIBYLL 2.3c. Small differences in the amount of energy violation between the two software packages are present, caused by CORSIKA's energy cutoff in tracked particles.  QGSJetII-04, SIBYLL 2.3c and UrQMD show a similar behaviour after the first interaction, violating approximately $\pm 5$\,GeV from the initial proton energy. In EPOS-LHC's case, events violating up to $\pm 15$\,GeV were registered. Currently, these anomalies are under investigation in collaboration with the models' authors.

To further constrain the initial interaction scenarios we define three inelasticity bands:  elastic and diffractive ($\kappa_1\approx[0,0.2]$), intermediate-inelastic ($\kappa_2 \approx[0.2,0.4]$) and highly inelastic collisions ($\kappa_3 \approx[0.4,1]$).
\begin{itemize}
    \item The $\kappa_1$ regime encompasses events with leading particles carrying energies above the threshold defined by the transition from high to low energy  hadronic model ($E^{\text{LP}}_{\kappa_{1}} =[80 - E_{p\text{,FI}}]$ GeV). These showers will undergo at least one additional interaction ruled by the high energy hadronic model, potentially leading to a larger impact in the ground level outcome. Furthermore, as most energy is concentrated in a single particle, the branching of the shower is very weak. We will refer as highly diffractive events to those in which $\kappa_{\text{Hel}}<0.05\subset \kappa_1$. This sub-regime represents events in which less than $5\%$ of the available energy is available for particle production in the first interaction. Besides, these showers are expected to replicate a scaled version of the first interaction spectrum in the upcoming interaction. 

    \item $\kappa_2$ showers have leading particles in the energy range $E^{\text{LP}}_{\kappa_{2}} =[60-80]$ GeV. The next interaction these showers undergo is governed by the low energy model, therefore the ground level observables will reflect the difference in the spectrum of particles that accompany the leading particle and their normalisation. 
    
    \item Finally, in the $\kappa_3$ region leading particles have energies in the range  $E^{\text{LP}}_{\kappa_{3}} =[0-60]$\,GeV. These events correspond to highly inelastic first interactions in which the model's initial particle production spectrum plays a determinant role.
\end{itemize}

\begin{figure}[hbt!!]
    \centering
   \includegraphics[width=1.0\columnwidth]{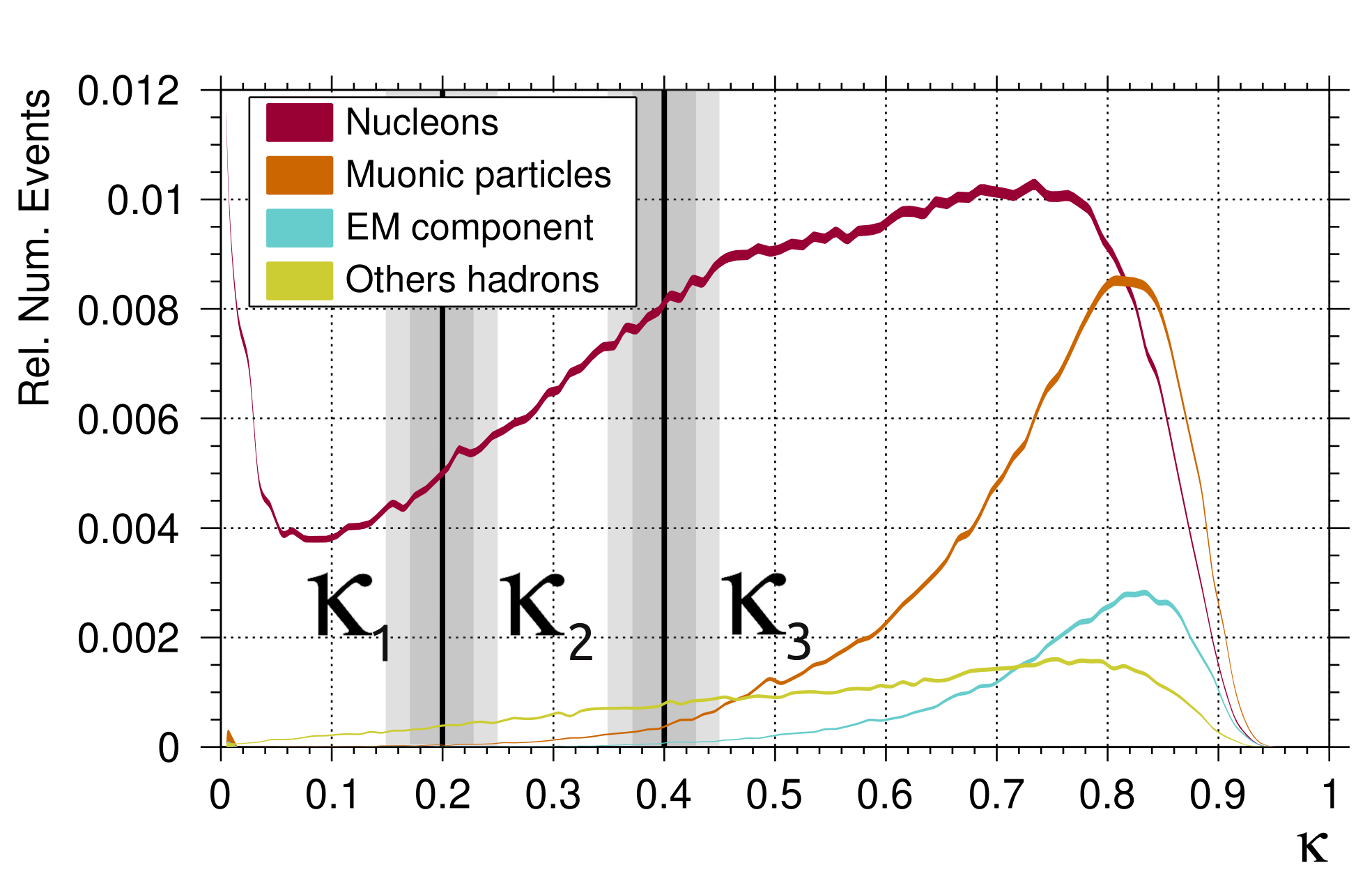}
    \caption{EPOS-LHC's inelasticity distributions for each leading particle family. The two shaded regions indicate 
    the systematic uncertainties in the definition of $\kappa$ due to event-by-event
    fluctuations in the amount of energy violation.}
    \label{EPOS_LP_ALL}
\end{figure}

In Figure \ref{EPOS_LP_ALL}, EPOS-LHC's  leading particle inelasticities are shown for each family. In most cases the leading particle is a nucleon, only at very large inelasticities particles from the muonic family dominate. Showers with leading EM particles or ``other hadrons'' contribute less than $7.5\%$ in all models. 

\begin{figure*}[htp!]
    \centering
    \includegraphics[width=1.0\textwidth]{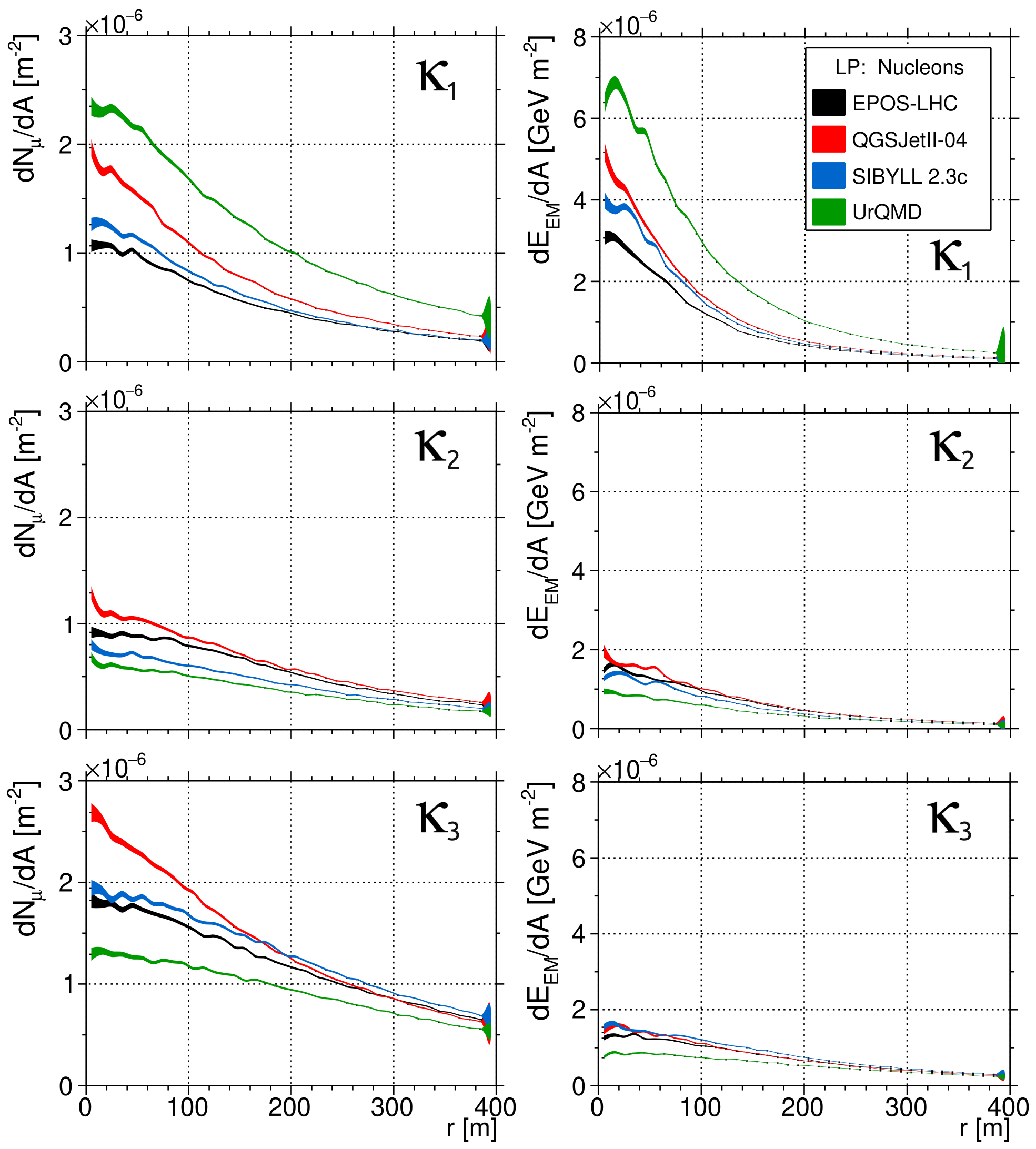}
    \caption{Breakdown of the muon (left panels) and EM component  (right panels) LDFs into the contributions from the three inelasticity regimes in nucleon led events. To have leading particles with similar properties, here energy ranges are taken as a proxy for the respective $\kappa$-regions, the boundaries of which are smeared due to event-by-event fluctuations in the energy violation of the models.}
    \label{MUEMLDF_kiprotons_divided_abs_rel}
\end{figure*}

\subsection{Lateral distribution functions} \label{LDFinelasticitybands}

To investigate the source of the disagreements between the models we study the ground level outcome arising from events with similar first interactions, focusing on the two leading contributions as explained in Appendix \ref{LDFAppSection}. In Figure \ref{MUEMLDF_kiprotons_divided_abs_rel} the four models' muon and EM  LDFs in the three classes are presented for events with a leading nucleon. As each $\kappa$-regime encloses broadly the same varieties of first interactions, we expect to obtain similar LDFs in each  contribution.  

The largest differences between the muon LDFs appear in the largest contributors, the $\kappa_1$ and $\kappa_3$ regimes. Initial collisions with large $\kappa$ produce abundant particle content (mostly muonic family particles) along with the leading particle, which later decay and contribute to the number of muons at ground level.  In events where the initial interaction is diffractive, higher inelasticity interactions occur deeper in the atmosphere, where the production of secondary particles from the muonic family will more likely affect the outcome at ground level. For the three high energy models, the largest ground level muon source originates from $\kappa_3$ events (highly inelastic interactions) while for UrQMD it is from $\kappa_1$ (diffractive interactions). 

In the $\kappa_3$ regime, the LDFs provide interesting information on the particle spectrum in the first interaction. Considering the integral over the LDF, SIBYLL 2.3c produces most muons ($1.85$ per this type of shower) followed by EPOS-LHC ($1.79$), QGSJetII-04 ($1.70$) and lastly UrQMD ($1.45$). The relative difference between SIBYLL 2.3c's and EPOS-LHC's average muon number is clear from the LDFs shape. As also seen in \cite{Sys_Diff_HIM}, QGSJetII-04 concentrates its smaller average number of muons at low core distances due to a lower muon transverse momentum  (to be discussed in the next section). 

In the LDF of the EM component, the largest contribution to the energy flow is present in the $\kappa_1$ nucleon led events. As in the muon LDF, a strong contribution from UrQMD's diffractive peak can be seen. A large contribution from $\kappa_2$ showers is not expected as the number of events in this regime is significantly lower $\left(\left. \frac{\text{N}_{\kappa_{2}}}{\text{N}_{\kappa_{3}}} \right\vert_{\text{EPOS-LHC}}\approx 0.29\right)$,
but also in highly inelastic events the ground level contribution to the EM component is small. This leads to the conclusion that the ground level EM component is driven by diffractive interactions. Physically this effect is understood, as at the energy of this study an EM component produced high in the atmosphere is very likely to die out before reaching ground level, while lower $\kappa$ interactions produce showers that develop deeper in the atmosphere and hence produce EM showers more likely to reach the ground. 

\begin{figure}[hbt!]
    \centering
    \includegraphics[width=\columnwidth]{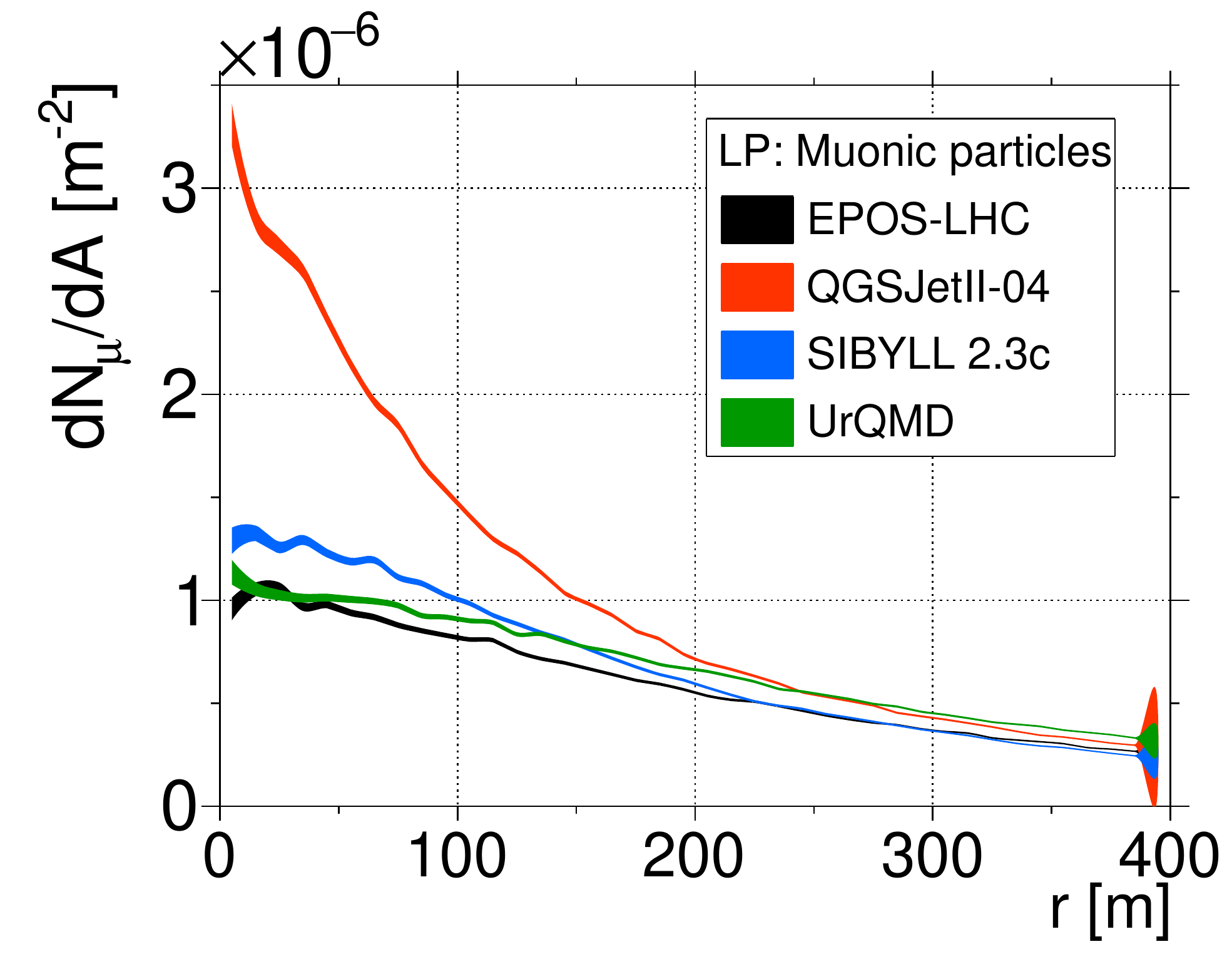}
    \caption{Muon LDF contribution from events lead by muonic family particles.}
    \label{MULDF_LPPions}
\end{figure}

Separately, in Figure \ref{MULDF_LPPions} the muon LDF from events with a leading particle from the muonic family  is presented integrated over all $\kappa$ ranges. In such events, a large fraction of the shower energy is assigned to multiple particles from the muonic family in the first interactions, therefore the large  differences are expected to be caused by  dissimilarities in the production of this type of secondary particles. For QGSJetII-04 the muon production at  short distances (r $\lesssim$ 200\,m) is approximately three times larger than for the other models. Similarly to the case of the $\kappa_3$ nucleon led muon LDF, the integral over the LDF does not indicate a sizeable difference in the absolute number of muons reaching ground level. QGSJetII-04's absolute number of muons from this type of scenario is only $4\%$ larger than UrQMD's, and $20\%$ more than SIBYLL 2.3c's and EPOS-LHC's. The observed excess at distances relevant for experimental purposes is therefore not expected to come from a larger number of events but from a different character of the muonic family particles.  

Concluding this section, we have seen that a breakdown of the ground level particle distributions into contributions from different types of initial interactions reveals striking differences between the models considered in this study. As the inelasticity ranges represent roughly similar physical events, a better agreement was expected.

\newpage

\section{First interaction analysis and sources of disagreement unravelled}

\subsection{First interaction rates}

\begin{figure*}[htb!]
    \centering
        \includegraphics[width=\textwidth]{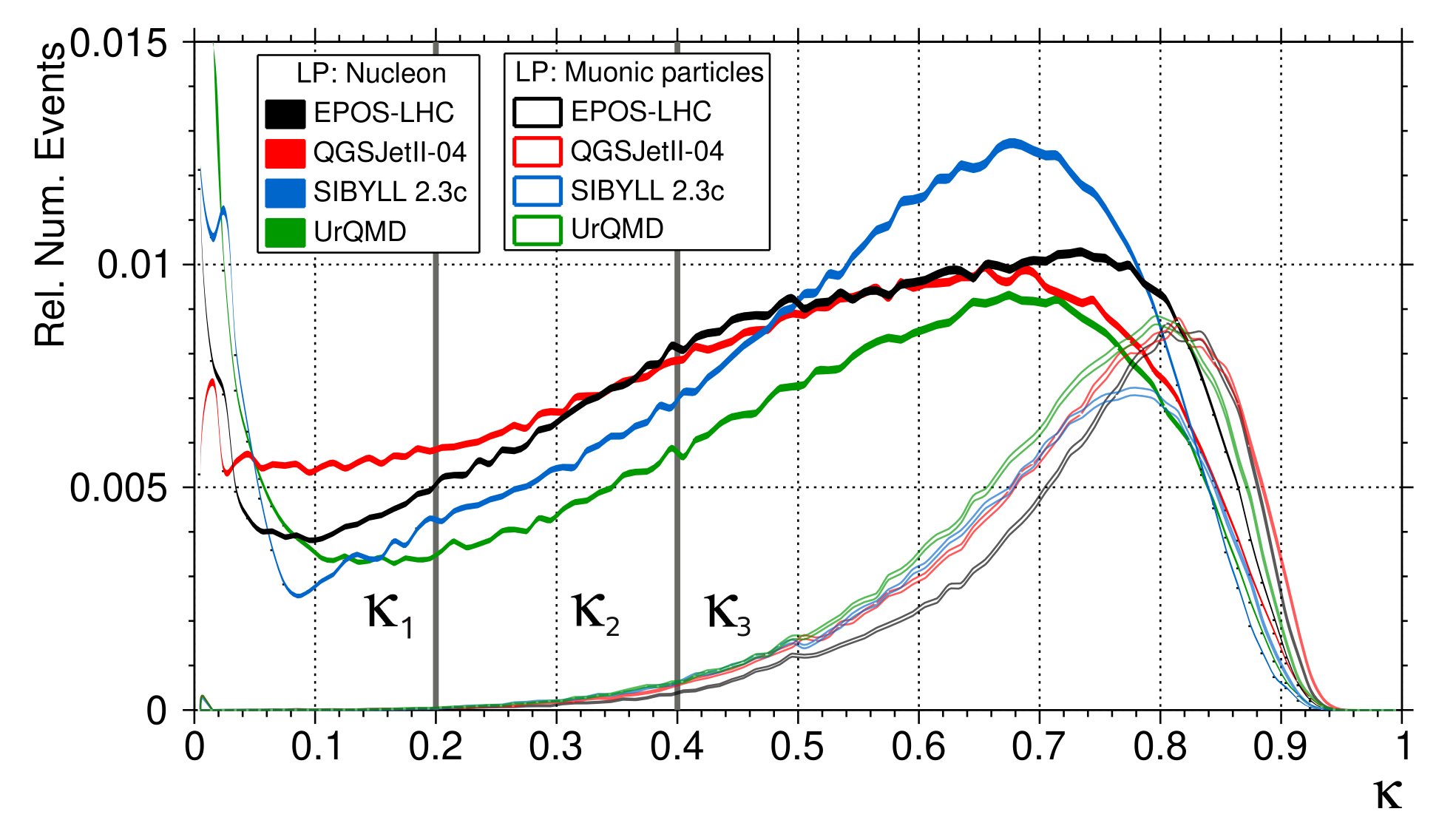}
    \caption{ Inelasticity distributions for the four studied models, for events initiated by the two dominant families; nucleons and muonic particles.}
    \label{LP_proton_pion}
\end{figure*}

In order to find the cause of the observed ground level differences we continue by studying the phenomenology of the first interaction. To understand the source of dissimilarities in the different inelasticity bands we begin  by analysing the  various leading particle inelasticity distributions for the two most contributing families, shown on Figure \ref{LP_proton_pion}.

Within the diffractive regime, UrQMD's number of events falling into the diffractive peak ($\kappa<0.05$) composes over $21\%$ of the total number of showers ($67\%$ of the $\kappa_1$ events and over five times EPOS-LHC's diffractive number). This abundance is responsible for the observed excess in both $\kappa_1$ LDFs (Figure \ref{MUEMLDF_kiprotons_divided_abs_rel}). As the shower development is spatially delayed, lower energy muonic family particles produced in secondary interactions will be more likely to contribute to the ground level products.  Additionally, the EM component is directly enhanced as the first interaction is only weakly inelastic and therefore, the EM shower maximum will be reached deeper in the atmosphere. Compared to the other models, UrQMD compensates for the high probability of low inelasticity events (Figure \ref{LP_proton_pion})  with a larger cross-section (Table \ref{crosssections} in Appendix \ref{Cross-sectionSection}), i.e. more interactions take place in UrQMD showers but they produce fewer particles on average.

Approximately $5\%$ of the total number of EPOS-LHC's and SIBYLL 2.3c's nucleon led events fall in the highly diffractive peak ($\kappa < 0.05$). However, it is necessary to recall the consequences of energy violation in the first interaction. In EPOS-LHC's case, the number of highly diffractive events significantly decreases if the first interaction is not considered to violate energy conservation. It implies that highly diffractive events lose considerable amounts of the shower energy and therefore, the leading particle energy is lower.

For QGSJetII-04 the $\kappa_1$ regime shows a difference in behaviour when compared to the other models. Its diffractive peak is localised and composed of only a small fraction of the total number of events. While the rest of the models show a minimum around $\kappa\sim0.1$, QGSJetII-04 maintains a constant level as soon as energy for particle production is available. This means that QGSJetII-04 has a higher rate of events with particle production than EPOS-LHC and SIBYLL 2.3c in the $\kappa_1$-regime. 
Such behaviour accounts for the absolute number of muons and EM component energy domination shown on Figure \ref{MUEMLDF_kiprotons_divided_abs_rel} (top panels). 
Furthermore, the dissimilar fractions of events at $\kappa \sim 0.1-0.2$ suggest an important source of disagreement for the event rate and particle production in primaries below the TeV scale. At higher energies a better agreement and smoother transition in this regime is reached, significantly reducing the relative differences in the ground level $\kappa_1$ contributions.

In the $\kappa_2$ regime, the distribution shapes broadly agree between models and only differ by a normalisation factor. Correspondingly, it is also shown in the middle panels of Figure \ref{MUEMLDF_kiprotons_divided_abs_rel}, where the larger number of events results in a larger number of ground level particles rather than in a change in the shape of the LDF. These events exhibit the best average ground level agreement as the low energy model governs most of the shower behaviour. The accompanying particle spectra are determined by the available energy, the leading particles of the different models place the shower development all under the same low energy model description. With increasing inelasticity more of the particle production happens in the first interaction within the high energy model, thereby again reducing the importance of the low-energy model for the description of the ground-level observables.

The $\kappa_3$ nucleon led distributions present large differences in the normalisation of event rates. For example, SIBYLL 2.3c strongly enhances large $\kappa$ collisions led by nucleons over muonic family particles, presenting no inelasticity regimes where the most likely leading particle is a muonic particle. 
The large variations seen in the $\kappa_3$ distributions show how the models' relative rates correlate with the observed differences in the muon number density in the $\kappa_3$ LDFs. It follows that even in inelastic events, the ground level differences arise from incompatibilities in same first interaction scenarios.

To close up this section, we have seen that the number of events falling in each inelasticity regime correlates with the hierarchy of the interaction products at ground level observed in the LDFs of Section \ref{LDFinelasticitybands}. This correspondence shows that a large part of the ground level differences originates from the models' dissimilar event rates. In other words, the normalisation differences emerge from a dissimilar nucleon-nucleon cross-section. Although certain  compensation between scenarios is reasonable (eg. $\kappa_3$ nucleon led and muonic family particle led), the strong disagreement around $\kappa\sim0.1$ reflects intrinsically large model incompatibilities.

\subsection{Accompanying particle production}
\begin{figure}[htb!]
    \centering
        \includegraphics[width=\columnwidth]{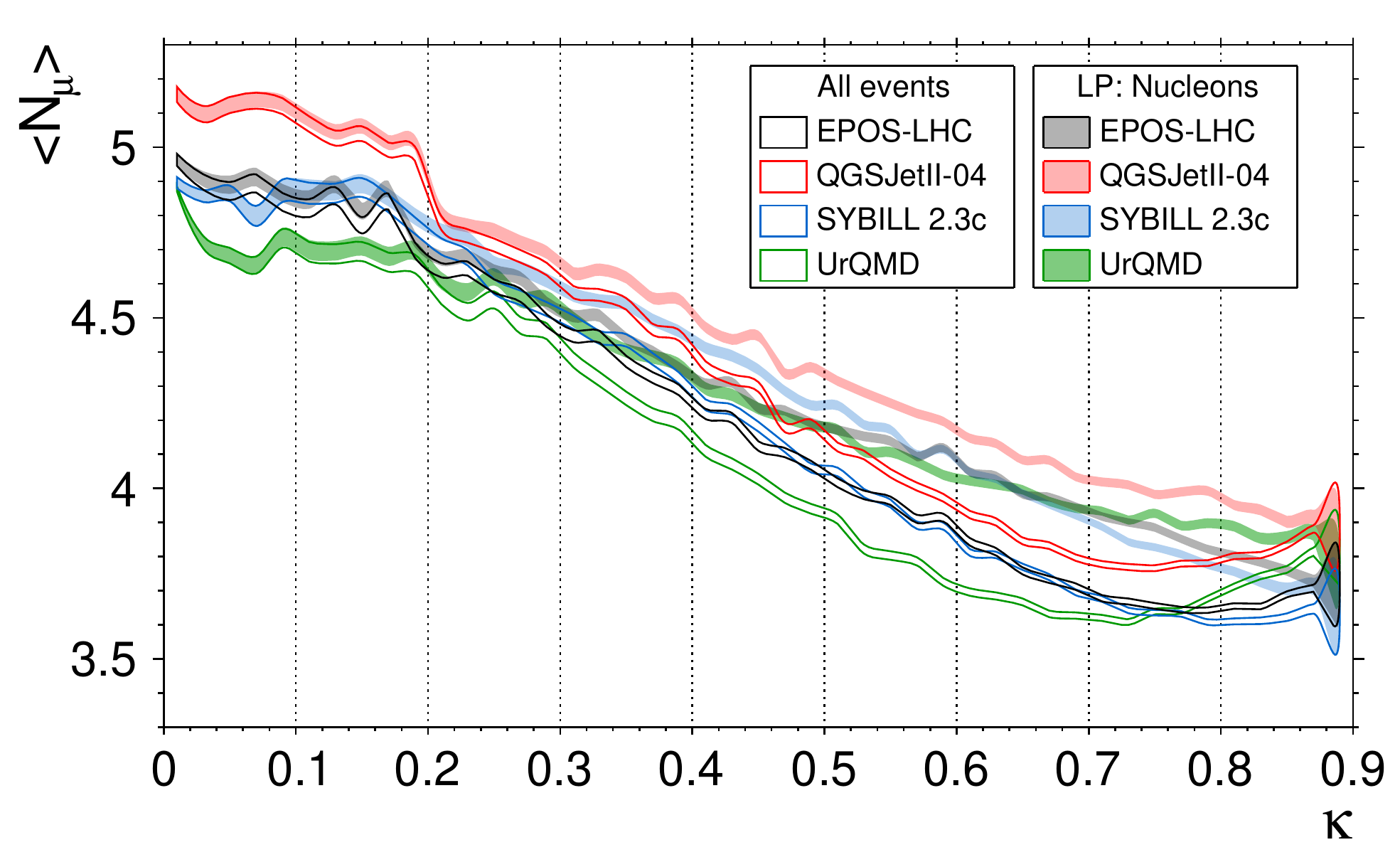}  
    \caption{Average number of ground level muons per event as a function of the first interaction inelasticity for nucleon led and all types of events.}
    \label{GLmu_vs_KLPprotons}
\end{figure}

In addition to the differences in the inelasticity distributions and consequently in the event rates, other characteristics have been found to directly affect the ground level muon LDF. Figure \ref{GLmu_vs_KLPprotons} shows the average ground level muon number per event as a function of the first interaction inelasticity for all types and nucleon led events. Regardless of the inelasticity distributions, some of the models generate a noticeably different number of muons at ground level per event.
For example, QGSJetII-04's events produce the largest average number of ground level muons for all events, but also on the selection of nucleon led events. The greater average number of ground level muons combined with the enhancement of the rate of events results in an overall larger muon number. Comparing both curves, it can be observed that although EPOS-LHC's and SIBYLL 2.3c's muon production in nucleon led events show  slight differences, they vanish when all events are considered. This illustrates how these two models have a similar ground level impact from a compensation between both leading families contributions. 
Furthermore, the larger number of $\kappa_1$ muons exhibits two interesting points; a stronger influence of the high energy model results in a larger number of muons, and secondly, a disagreement between the high and low energy models. 

Due to the transition of the high energy model to the low energy model, there is a reduction of muon number at $\kappa\sim0.2$, above which all secondaries from the first interaction are handled by the low-energy model. The drop is most pronounced for QGSJetII-04. This illuminates the general difference between high and low energy models and the impact of the secondary interactions. As we are plotting a ground level quantity in terms of an first interaction parameter, the distributions shown are sensitive to the late shower development. At these inelasticities, the influence of the low energy model on the shower is at its largest as the transition energy was set to 80\,GeV and the leading particle's next interaction will be determined by the UrQMD model. QGSJetII-04's sudden change in slope shows how events in which the leading particle is ruled by the low energy model produce a lower number of muons at ground level (as UrQMD overall does). Such convergence towards UrQMD's nature shows how showers are built from dissimilar model approaches and how sensitive the ground level number of muons is to the model switch.

In order to explain the general muon excess from QGSJetII-04 (compared to the other models), it is  not enough to only look at the leading particle in the first interaction, but also the accompanying particles need to be studied. The early production of particles from the muonic family will directly impact the ground level muons.  
The minimum energy for 50\% of the muons to travel approximately $17$\,km without decaying is $\sim5$\,GeV. We will use this energy as a threshold to select the first interaction muonic family particles that contribute directly to the ground level muon number.

\begin{figure}[htb!]
    \centering
    \includegraphics[width=\columnwidth]{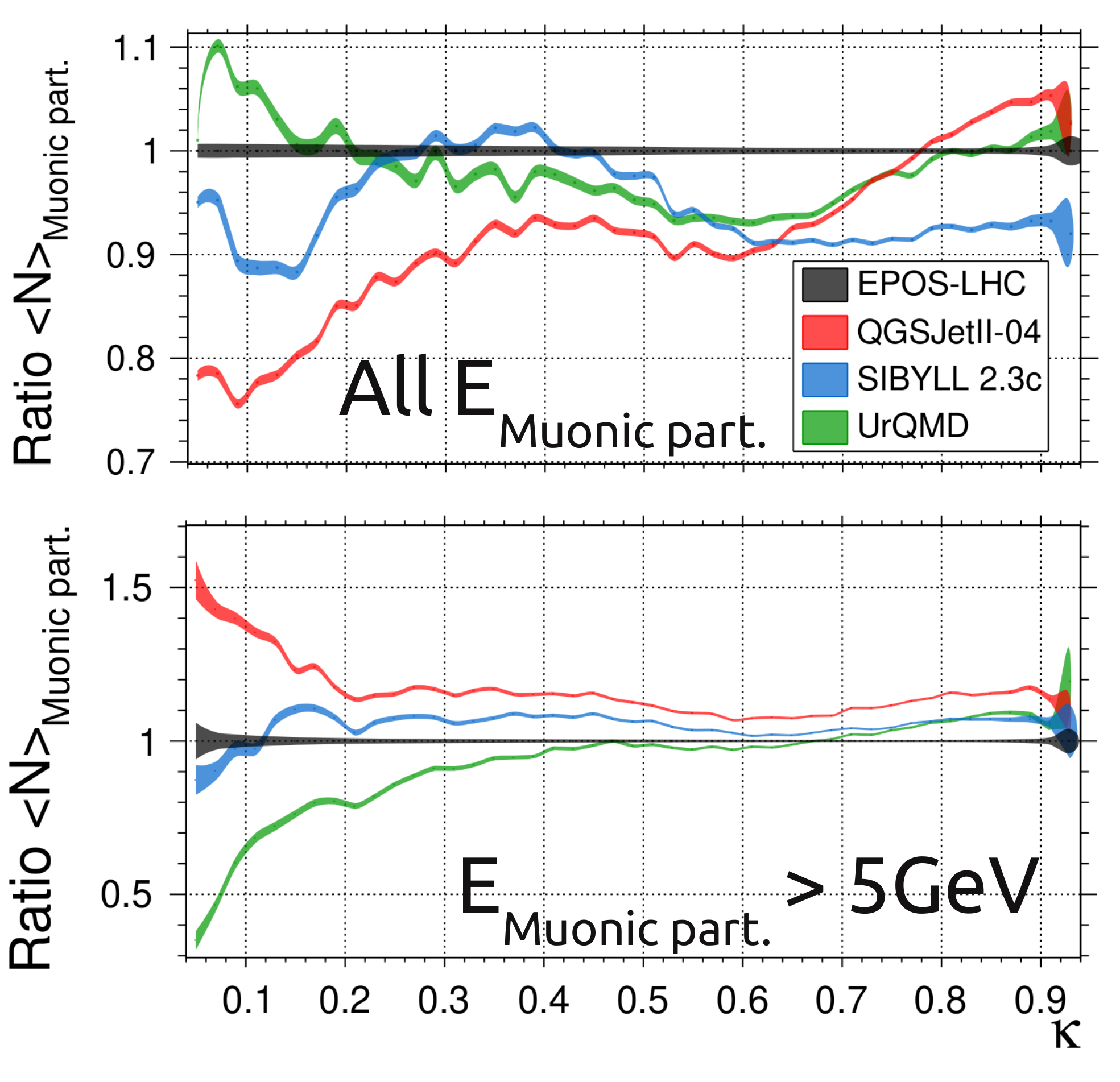}
    \caption{Muonic family particle production ratios with respect to EPOS-LHC considering all particle energies (top panel) and above an energy threshold of 5\,GeV (bottom panel).
    }
    \label{averageNpions_PFI_vsK}
\end{figure}

In Figure \ref{averageNpions_PFI_vsK}, the average numbers of muonic family particles produced in nucleon led events are studied. On the top panel, the ratio between the average number of all accompanying muonic particles is presented and in the lower panel, the cut of 5\,GeV is applied. Considering all muonic family particles produced, QGSJetII-04's average larger ground level muon number is not evident as it does not show an increased abundance over the other models. However, when considering high energy muons, QGSJetII-04 exhibits an average effective muonic particle surplus over the other models, accounting for the average shower muon excess at ground level. 

The investigation in \cite{Sys_Diff_HIM} of the muonic family particle production and ground level muon number at higher primary energies shows that a better agreement between the models is not found by increasing the primary energy. This means that the discontinuity in the description of hadronic physics when switching from the high-energy to the low-energy model is not reduced. A consistent description of particle production would require a well defined transition energy where the physics implemented in the models agrees.

\subsection{Transverse momenta of first interaction muonic family particles and  distribution of ground level muons}
\begin{figure}[hbt!]
    \centering
    \includegraphics[width=1\columnwidth]{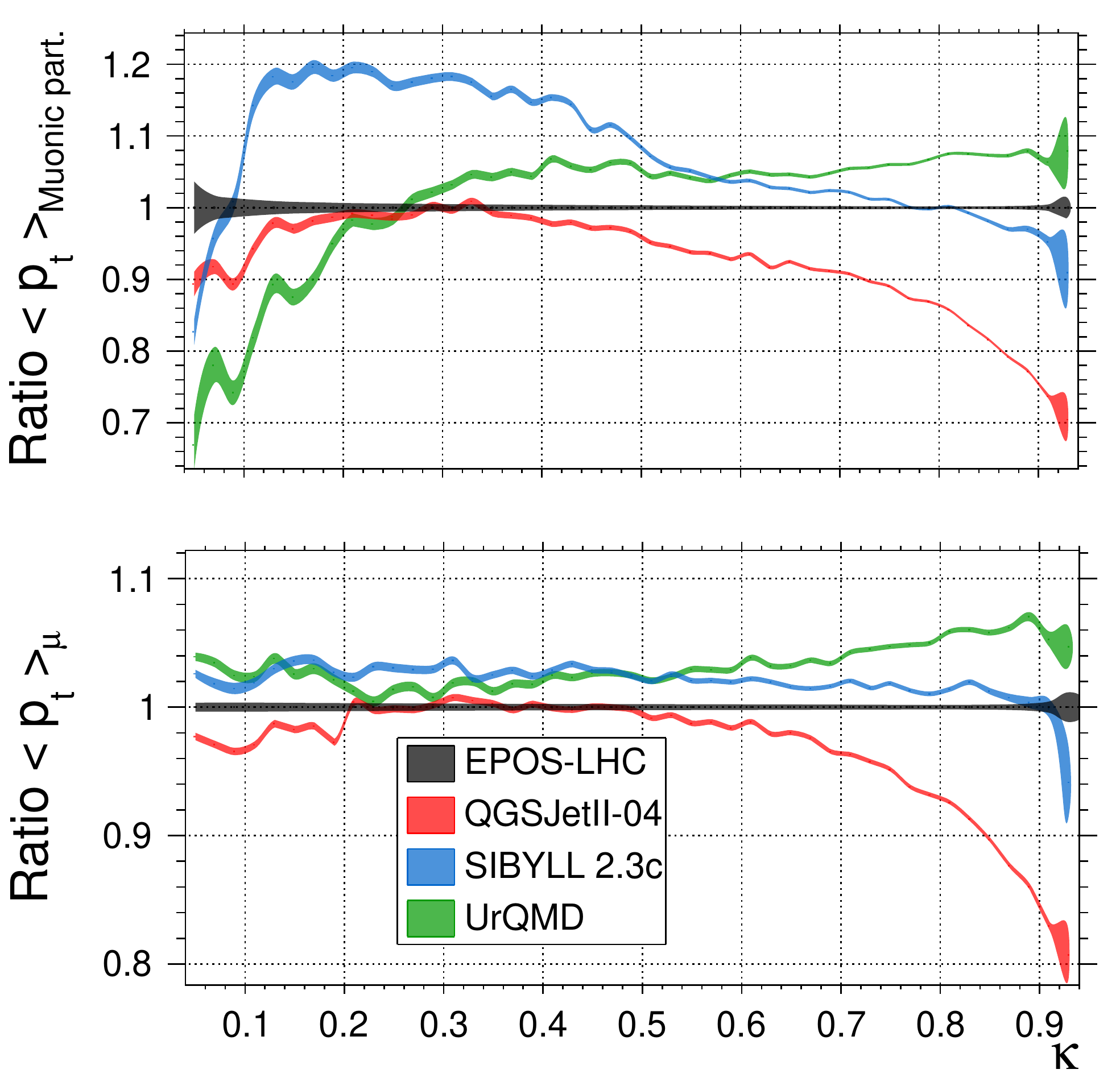}
    \caption{ Ratios with respect to EPOS-LHC of the (top panel) first interaction muonic family particles (energy above 5\,GeV) and (bottom panel) ground level muons average transverse momenta as a function of the first interaction inelasticity.}
    \label{MM_muons_pt_K}
\end{figure}

As discussed in section~\ref{sec:FirstInteraction},  the number of muons at small distances (r$<200$\,m)  from the shower core produced in events with large inelasticity ($\kappa_3$) and with a leading nucleon or particle from the muonic family does not reflect the total number of produced muons. The difference seen can instead be understood from the larger number of events falling in this regime and the average muonic family particle production. An explanation for the muon concentration (QGSJetII-04's case) and a spread (UrQMD's case) is important for ground-level observations. A parameter directly related to this spread is the transverse momentum of the ground level particles; lower $p_{\text{t}}$ muons concentrate at short distances while large  $p_{\text{t}}$ are spread away from the shower axis. In Figure \ref{MM_muons_pt_K} the mean transverses momentum from energetic muonic family particles in the first interaction and ground level muons is shown. From the top panel it can be concluded that as inelasticity increases and more energy is being assigned to muonic particles, the transverse momenta significantly diverge between models. This disagreement in the first interaction is directly correlated to the muon $p_{\text{t}}$ at ground level, resulting in the discussed differences from Figures \ref{MUEMLDF_kiprotons_divided_abs_rel}  (bottom left panel) and \ref{MULDF_LPPions}. The large differences between models in the transverse momentum spectra of muonic family particles (mostly produced in large $\kappa$ events) and their extrapolation to ground level muons was already pointed out in \cite{Sys_Diff_HIM}, where the higher energies muonic particles were shown to diverge between models. 

Worthy of note is the slight reduction of transverse momentum of ground level muons in low inelasticity collisions and the abrupt change at $\kappa=0.2$. As was commented in Figure \ref{GLmu_vs_KLPprotons}, this is caused by the switching from high to low energy models. This is most clearly seen between models with most different muonic family particles production and properties, eg.  QGSJetII-04 and UrQMD. 

\section{Model Summary}\label{ModelSummarySection}
In this section, the main characteristics and differences of the models are summarised:
\begin{itemize}
    \item EPOS-LHC ground level components present a relative deficit in the contribution from $\kappa_1$ events, as shown in the muon and the EM component LDFs. This deficit is a consequence of the weak impact of EPOS-LHC's diffractive collisions as their leading particles are not as energetic as in the other models. Moreover, this model is more affected by the energy violation in the first interaction. Over 10\,GeV variations were registered which influenced the computation of the first interaction inelasticity. The diffractive peak at $\kappa<0.05$ would be completely smeared out, as the leading particles from highly diffractive events have on average lower energies in EPOS-LHC than in the other models.

    Moreover, EPOS-LHC's production of accompanying high energy muonic family particles is slightly lower than in the other models but compensated by a larger production of secondary high energy nucleons.
    
    \item QGSJetII-04 has been shown to present a significantly larger number of muons at ground level. Moreover, the excess is critical 
    at short core distances where simulations are most relevant for experimental purposes.
    
    The overproduction arises from  a different energy spectrum of secondary particles that does not  cause an overall larger number of muonic family particles but an increase in the amount of those carrying energies over 5\,GeV. Additionally, highly diffractive events are suppressed while at larger inelasticities, where more energy is available for particle production, the rates are enhanced. 
    
    QGSJetII-04's muon concentration at short core distances is caused by lower transverse momenta in the muonic family particles produced. It is clearest at large $\kappa$ values where muonic particles carry great fractions of the shower energy. The muonic family particles' $p_{\text{t}}$ is transferred in their decay to the muons which reach ground level closer to the shower axis.
 
     \item SIBYLL 2.3c's ground level spectrum is similar to EPOS-LHC's. Despite this agreement, both models have been shown to be very different; the event rate at large $\kappa$ values is significantly different as well as the accompanying particle production. In the $\kappa>0.4$ regime, events led by nucleons with a strong accompaniment of particles from the muonic family are enhanced over events that are led directly by muonic particles. In contrast, EPOS-LHC's $\kappa_3$-nucleon led events have a larger fraction of accompanying nucleons but result in a subtle difference at ground level. 
    
    \item UrQMD's proton cross-section is shown to be significantly larger than that of the other high energy models in Appendix \ref{Cross-sectionSection}. This has a severe impact on the development of showers initiated by 100\,GeV primaries if the height of the first interaction is not fixed. Furthermore, we  have shown that over 20\% of the events are highly diffractive, showing a great difference with the other models. Although a compensation between the two effects could be expected in non-fixed first interactions, it has been shown in \cite{Sys_Diff_HIM} that this does not occur and causes an overall deficit of the ground level components.
    
    In comparison to the high energy models and specifically to QGSJetII-04, UrQMD's high energy muonic family particles show large transverse momenta. Consequently, its muon LDF is  spread over a larger surface, resulting in the opposite behaviour to QGSJetII-04. 
    
    UrQMD's observed disagreement in the production of muonic particles ($>5$\,GeV) and their transverse momentum, questions the setting of the transition energy. It has been shown that in the $\kappa_2$ regime all high energy model predictions blend together due to the large influence of UrQMD's physics in secondary interactions. In contrast, all models seem to diverge in regimes where the low energy model does not dominate; this can be seen in $\kappa<0.2$ and $\kappa >0.5$ events.  
   
\end{itemize}

\section{Impact of model choice and transition energy setting on air shower development}

\begin{figure}
    \centering
    \includegraphics[width=1\columnwidth]{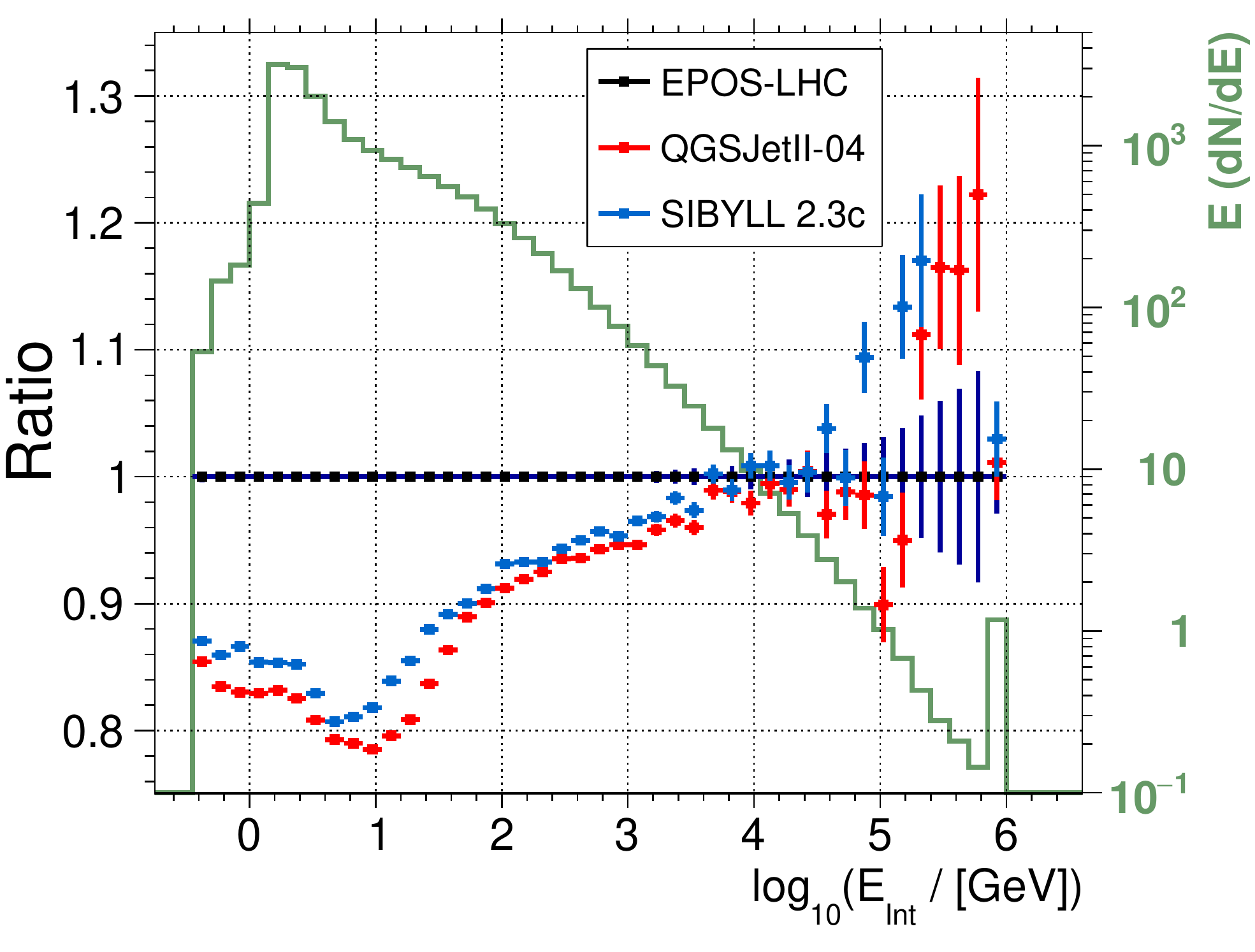}

    \caption{On the left y-axis, the ratio between the hadronic interaction distributions in PeV proton showers is shown for each high energy model with respect to EPOS-LHC. Labelled in green and on the right y-axis, EPOS-LHC's hadronic interactions distribution is shown for a PeV proton initiated event.
    }
    \label{DiffswEHEmodels}
\end{figure}

So far, the main focus has been on the interaction of 100\,GeV protons with atmospheric nitrogen. An air shower can be viewed as a superposition of cascades initiated by particles with a fraction of the initial particle energy. Therefore, it is expected that the model differences observed for interactions with 100 GeV primaries, might also manifest themselves throughout the development of the air showers that were initiated by primaries with energies well above 100\,GeV. Moreover, as the event generators select the hadronic interactions by the centre of mass energy and the participants involved, the correspondence between the studied first interactions and the mid-shower development is ensured.  It is not practical to track the impact of each individual sub-shower generated by a hadronic interaction on the air shower as whole. However, to get an idea of the impact of the hadronic interaction model choice and the transition energy between low and high energy models, the CORSIKA data access tools (COAST) package \cite{corsika} was used to track the rate of hadronic interactions throughout the whole shower development.  This analysis contextualises the studied differences at higher energies and more commonly studied primaries. Shifting the transition energy therefore reveals whether the studied differences enhance the disagreement or bring models into accordance.

In Figure \ref{DiffswEHEmodels} we  display the hadronic interaction ratios in PeV proton initiated showers (with default transition energy). Firstly, EPOS-LHC's events undergo a considerably larger number of interactions below the 10\,TeV scale, over 10\%  more than QGSJetII-04 and SIBYLL 2.3c at energies at which muon production peaks. This reflects how, for the same initial conditions, models generate different shower scenarios which agree in their ground level product \cite{Sys_Diff_HIM}.  Consequently, the differences in particle production and shower dynamics between models can also be inferred here. Worthy of note is QGSJetII-04's and SIBYLL 2.3c's high energy interaction rate. Although the  likelihood of these collisions  is well suppressed, the whole shower is affected by a larger number of interactions carrying large fractions of the initial energy. 

We show in the Appendix \ref{TransEHadProf} the effect of modifying the values of the transition energy on the hadronic interaction profiles. There it is shown how the increase or decrease of the transition energy results in opposite effects for EPOS-LHC and the other models. Strengthening the role of the low energy model causes a decrease in the number of interactions under 10\,TeV in EPOS-LHC and an increase in QGSJetII-04's and SIBYLL 2.3c's. This can be understood as UrQMD's hadronic interaction profile lying in between EPOS-LHC's and SIBYLL 2.3c's. 

In the previous sections we discussed how the models, high or low energy, do not agree for the default transition energy value. While a better compatibility in the inelasticity distributions can be found at higher energy, increasing the transition energy could mute the properties of the individual high energy models and threaten the consistency of events at higher energy. This inconsistency and abrupt change in the secondary interaction physics demands for a better agreement between models and/or a redefinition of the transition scales.

\section{Conclusions}
Through a detailed investigation of events initiated by 100\,GeV protons we cast light on the sources of disagreement between hadronic interaction models \cite{Sys_Diff_HIM}. By a simultaneous analysis of the first interaction and ground level particles, we can conclude that models present an intrinsically different behaviour in the low energy regime. An outline on each model's arguments is presented in section \ref{ModelSummarySection}. 

Phenomenological variables were used to correlate differences in ground level observables to properties of the first interaction for some commonly employed high energy models. Discrepancies in the ground level particle number emerged from event rate deviations that were drawn from the study of first interaction inelasticity distributions. Within same events types (same leading particle and inelasticity), the production of accompanying particles (eg. muonic family particles) disagrees causing differences in the ground-level muon component. The transverse momentum spectra of muonic particles that are produced in the first interaction has also been shown to differ between models. This disagreement becomes worse as their energy increases.

The focus in this study was on the differences between models in the early shower development, however, the model differences are of course not restrained to the early development stage.  When increasing the primary energy, the number of interactions around the switching energy will increase, however, the differences on the average shower parameters are washed away by the dominance of particle production by the low energy model. When selecting air showers with special properties, like proton induced air showers that mimic gamma-ray induced air showers, differences between the modelling of
hadronic interactions might be exposed again \cite{CTA_HADR}, also at higher primary energies (TeV domain). 

Additionally, we have shown how the event generators diverge by studying the number of hadronic interactions in showers initiated at high energies. Large differences have been spotted remarking how the models produce very dissimilar showers from the same initial conditions. In this line, we have shown how sensitive the shower development is to modifications of the transition energies. The low energy model shows a clear domination of the shower (ruling over 85\% of the interactions in a PeV shower with default transition energy and increasing percentage for higher primaries) while the high energy models set the initial conditions. This points to low energy model investigations as a possible important contribution to the solution of the hadronic interaction model puzzles \cite{muondeficit,HansMuonPro,Dembinski:2019uta}.

As the focused energy regime falls within the validity range of all models, a reasonable agreement would be expected between all high energy models and UrQMD. The large differences found demand for a convergent model tuning to  existing and/or future accelerator data or a redefinition of the transition regime from high to low energy hadronic interaction models  that ensures consistency with the high energy partners.

\section{Acknowledgements}
 The authors thank the MPIK non-thermal astrophysics group for fruitful discussions about the paper, in particular J.A. Hinton for providing helpful comments.

\bibliographystyle{elsarticle-num_etal} 
\bibliography{./references_AP}

\appendix
\section{Proton-air cross-sections}\label{Cross-sectionSection}

\begin{table}[t]
\caption{First interaction results from $10^6$ event simulations with non-fixed initial altitude. Values are obtained for the atmospheric model MODATM 22 (GDAS/May). }
\begin{tabular}{|c|c|c|c|}

\hline 
{Model} & $\lambda_{\text{p-Air}}$ [g cm$^{-2}$] &\quad $\sigma$ [mb] \quad \quad  & First interaction\\ & & & altitude [km]  \\ \hline \hline
EPOS-LHC & 87.81 & 275.12 & 17.34  \\ \hline
QGSJetII-04 & 90.83  & 265.99 & 17.13  \\ \hline
SIBYLL 2.3c & 85.95  & 281.09 & 17.47  \\ \hline
UrQMD & 75.72  & 319.07  & 18.26  \\ \hline \hline
Average & 85.08 & 285.32 & 17.55  \\ \hline
\end{tabular}
\label{crosssections}
\end{table}

In the study of hadronic interaction event generators, the altitude at which the first interaction occurs is determinant for the shower development and the number of particles arriving to ground level. The altitude is dependent on the proton mean free path implemented in each cross-section and the atmospheric density model. As cross-sections are inherent to each model, they were computed at 100\,GeV to quantify one of the most important initial shower conditions differences. 

Simulations  of $10^6$ events with free first interaction altitude were performed for each model, such that statistical uncertainties are negligible. Following \cite{cross-sections_protonair}, the distribution of first interaction points ($X_0$) was fitted with
\begin{equation}\label{crosssectionFIpoint}
    \frac{\text{d}N}{\text{d}X_0}=\frac{1}{\lambda_{\text{p-Air}}}\exp{\left[-\frac{X_0}{\lambda_{\text{p-Air}}}\right]}
\end{equation}
from where the mean free path for protons in air was computed.  The mean free path and the air's average mass are related by
\begin{equation}
\sigma_{\text{p-Air}}=\frac{\langle m_{\text{Air}} \rangle}{\lambda_{\text{p-Air}}}
\end{equation}
and taking $\langle m_{\text{Air}} \rangle\approx14.45\,m_{\text{p}}$. Finally, using the atmospheric model density distribution with GDAS/May parameters, the respective altitudes were computed (displayed on Table \ref{crosssections}). 
 
Our estimations for the high energy models agree with the results in detailed cross-section studies and experimental data reviews, e.g. \cite{cross-sections_protonair}. The differences between the high energy models ($\sim10$\, mb with respect to EPOS-LHC) are well known and pointed out in the cited study. However, UrQMD presents a considerably larger total cross-section, causing a starting point higher about 1 km, on average. As a consequence, the shower development occurs higher in the atmosphere substantially affecting  ground-level behaviour of the shower. 
 
As we want to analyse the first interactions and additionally reduce the effect of UrQMD's high cross-section, the first interaction in the simulations was fixed at 17.55\,km.

 \section{Average particle lateral distributions}\label{LDFAppSection}
 
\begin{figure*}[htb!]
    \centering
    \includegraphics[width=\textwidth]{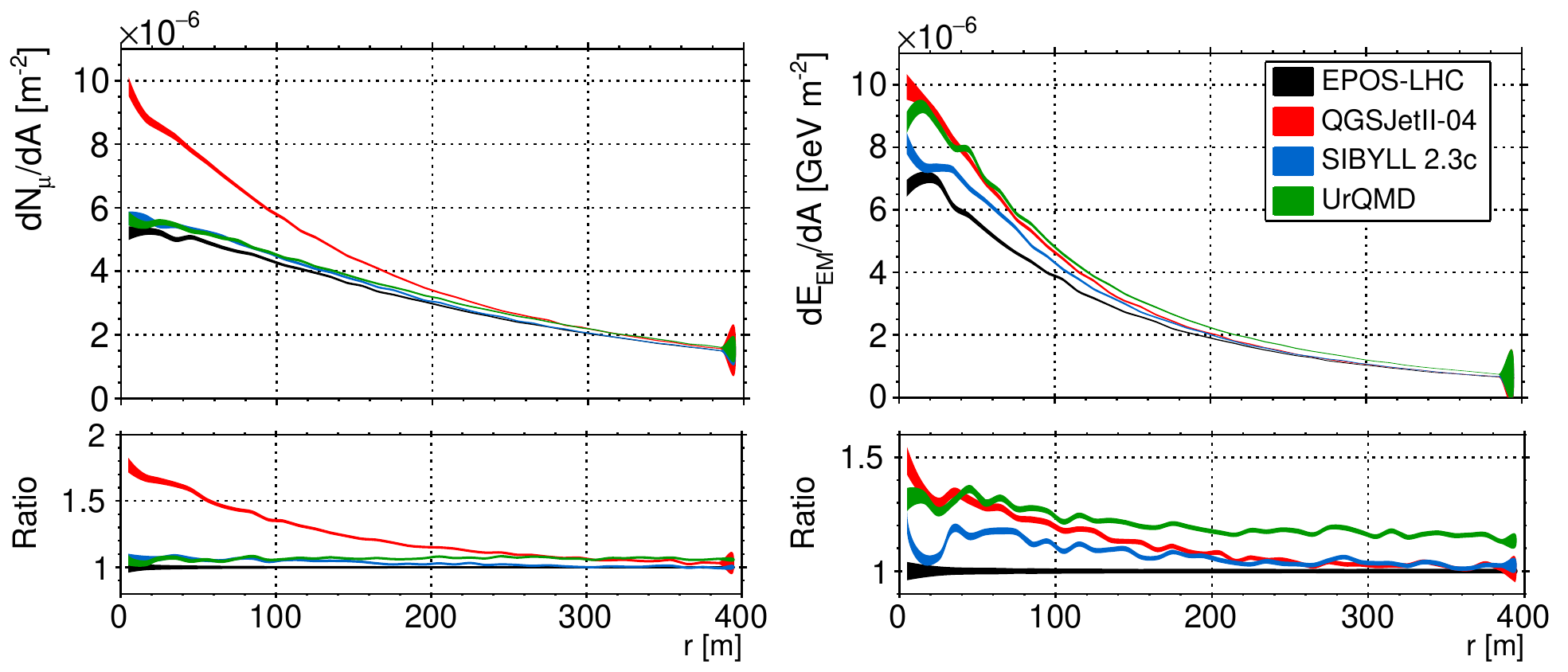}
    \caption{ (Top left panel) Average muon LDFs for the four  hadronic interaction models at ground level (4100\,m). (Bottom left panel) The LDFs ratios with respect to EPOS-LHC. (Right panels) Analogous to the left panels but for the the energy deposit in the EM particles at ground level.}
    \label{MUEM-BareLDF}
\end{figure*}

Figure \ref{MUEM-BareLDF} shows the muon and EM (combined electron and gamma-ray) component LDFs with the corresponding ratios between models. In the case of muons the LDF is shown as a particle count, while for the EM component it is shown as an energy deposit in an attempt to replicate the typical behaviour of ground-based particle detectors. Here we discuss the total ground level distributions while in the main text the breakdowns into particular  initial scenarios were analysed. 

Compared to EPOS-LHC, for the EM LDF the other models show an enhancement between $20-50\%$  in the central region with a radius up to 200m. On the other hand, the muon LDF of EPOS-LHC, SIBYLL 2.3c and UrQMD look rather similar. However, in this comparison QGSJetII-04 presents an extreme outlier, with almost 80\% higher muon density close to the shower core.

Comparing these LDFs in Figure \ref{MUEM-BareLDF} with those presented in previous studies \cite{Sys_Diff_HIM},  we observe agreement in the ratios for the high energy models. However, UrQMD's relative deviation with respect to EPOS-LHC observed here is different from  the one shown in \cite{Sys_Diff_HIM}. The muon number and EM energy excess in this research's simulations are given for a fixed height of $\sim17.5$\,km. As explained in the previous section, UrQMD's cross-section is considerably larger causing, in free first interaction simulations, an earlier initial interaction and hence a lower number of particles reaching the ground level. In \cite{Sys_Diff_HIM} the simulations were performed with a free first interaction point reflecting the effect of  UrQMD's large cross-section. 

The differences shown are in all cases found at small core distances, where simulations are most relevant for experimental purposes. By differentiating the ground level observables into the contributions from similar nature events, we have shown in Section \ref{LDFinelasticitybands} that the total EPOS-LHC's, SIBYLL 2.3c's and UrQMD's muon LDF agreement (Figure \ref{MUEM-BareLDF}) presented here arises as a compensation between the different regimes.

\begin{figure*}[hbt]
    \centering
    \includegraphics[width=\textwidth]{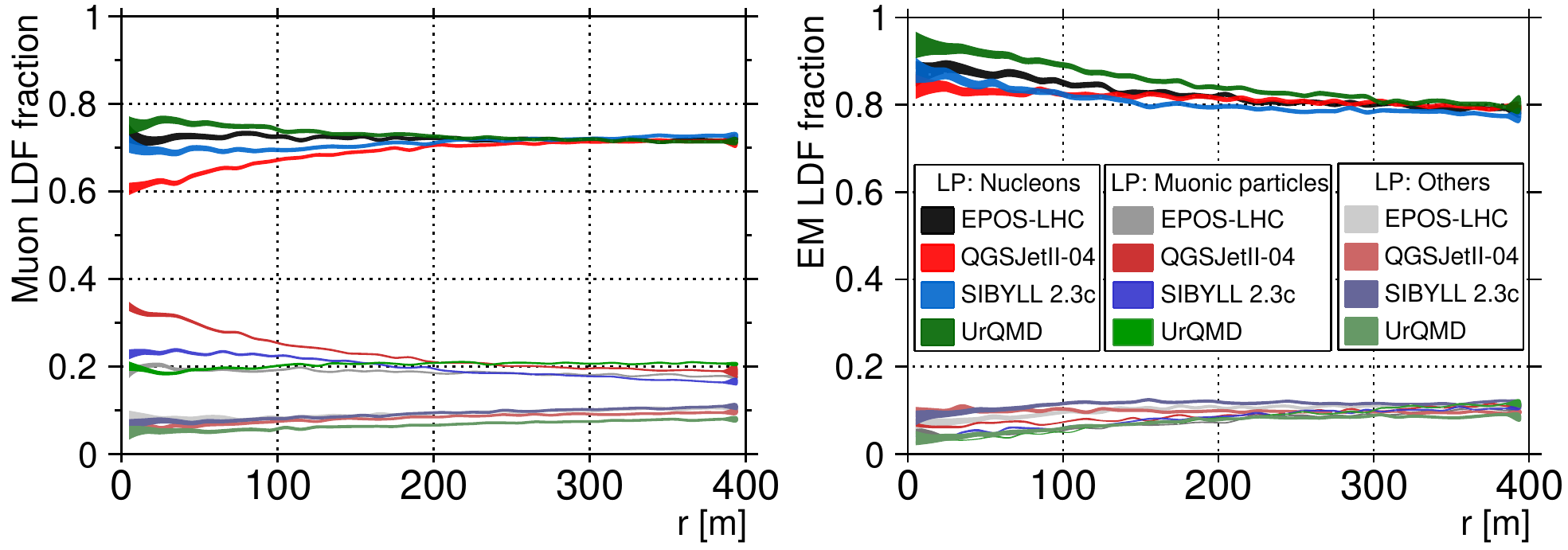}
    \caption{Fraction of the  muon (left) and EM (right) LDF contribution for different leading particle families in each hadronic interaction model.}
    \label{muemldf-prot-pi-fractions}
\end{figure*}

\subsection{Leading particle fractions}

In an attempt to identify the sources of ground level disagreements, the LDFs can be separated into the contributions arising from the leading particle family. Figure \ref{muemldf-prot-pi-fractions} shows the total LDF fraction arising from nucleon and muonic family particle led showers. As the contributions from EM component and ``other hadrons" particles led events is relatively smaller, they are united into a single group. 

In all models, most muons reaching ground level originate from nucleon led events, followed by muonic family particles. Only around $10\%$ of the events have a leading particle that is not a nucleon and does not belong to the muonic or EM family. The good agreement in EPOS-LHC's, SIBYLL 2.3c's and UrQMD's fractions together with the similar total LDFs show that the muon production is, in these models, independent of the leading particle type. 

QGSJetII-04's relative contributions are however significantly different. At short core distances, the fraction of muons arising from muonic particles led events is noticeably larger. This demonstrates the important role of this type of events for muon production within QGSJetII-04 model.  

Although the fraction of muons at short core distances arising from QGSJetII-04's nucleon led events is relatively smaller, the total number of muons created by these events and falling in $r<50$\,m is equal to the overall contribution from the other models. The larger muon production in both event classes and the dissimilarity in the contribution at short core distances suggest an overall different muon production in QGSJetII-04. 

Over $80\%$ of the EM energy arriving at ground level originates from nucleon led showers. Although QGSJetII-04's and UrQMD's total EM LDFs (shown more clearly on Figure \ref{MUEM-BareLDF}) appear to agree, the difference in the nucleon led showers contribution shows differences in leading particle contributions and hence the model physics. 

Recalling the cross-section discussion in Appendix \ref{Cross-sectionSection}, UrQMD's ground level LDF is strongly dependent on the first interaction point, therefore the arbitrary matching with QGSJetII-04's EM LDF is a consequence of the fixed altitude chosen for the first interaction. By leaving the first interaction point free, the relative difference between the high energy models nucleon led contributions would remain unchanged while UrQMD's nucleon led shower input would decrease as a result of its high cross-section.

\section{Transition energy effect on the hadronic interactions profile in PeV showers.}
\label{TransEHadProf}
\begin{figure}
    \centering
    \includegraphics[width=1\columnwidth]{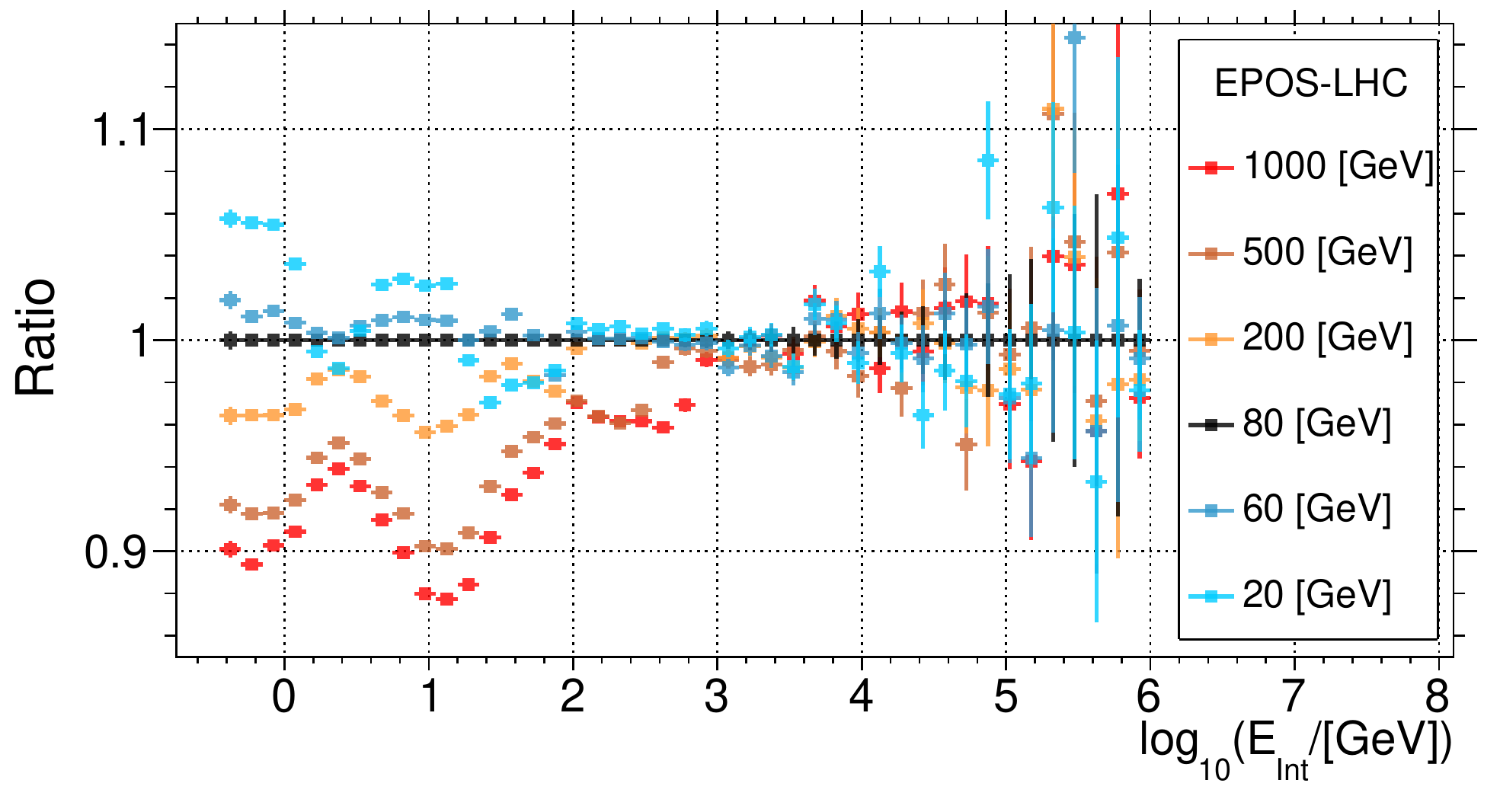}
    \includegraphics[width=1\columnwidth]{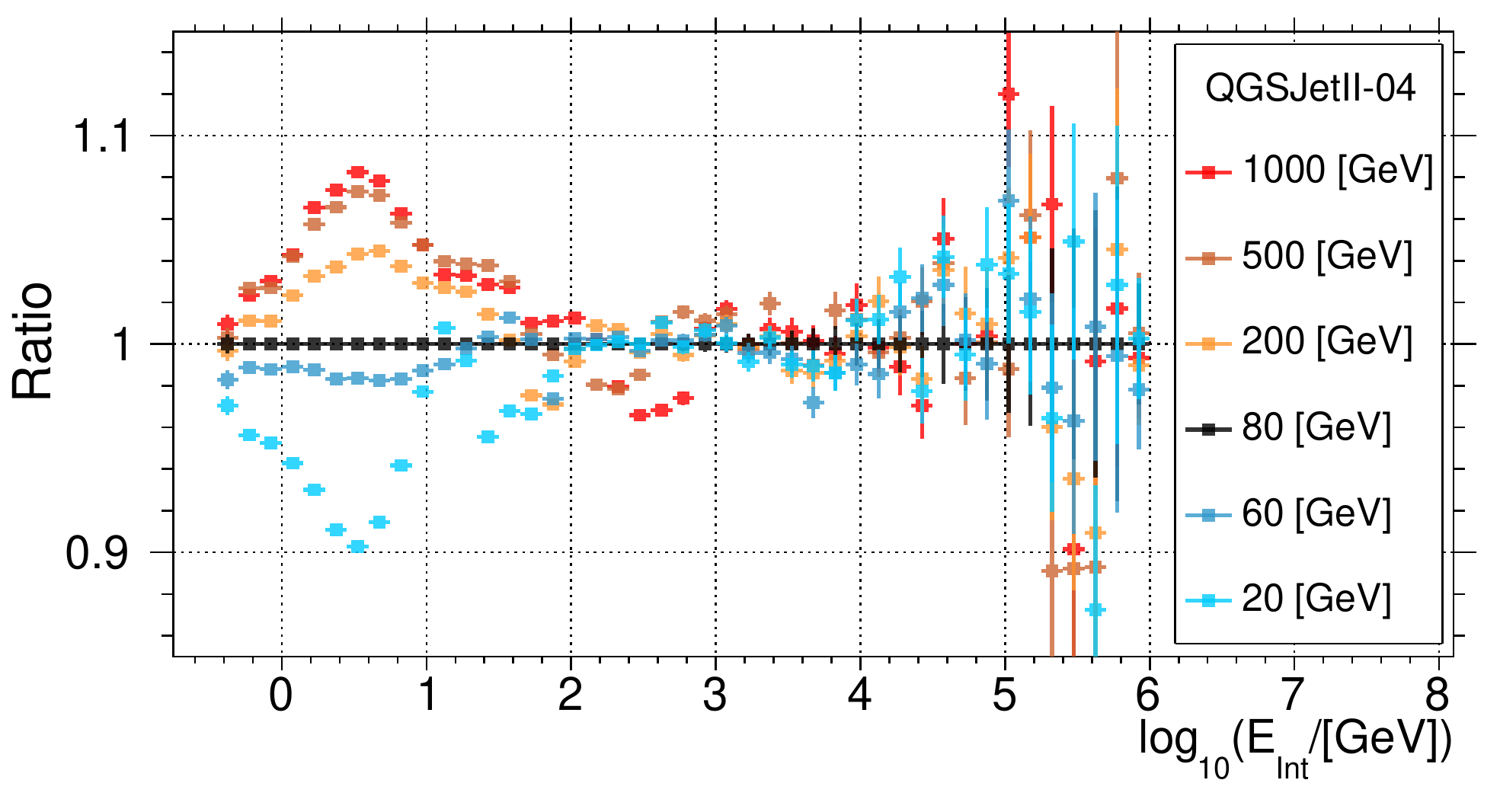}
    \includegraphics[width=1\columnwidth]{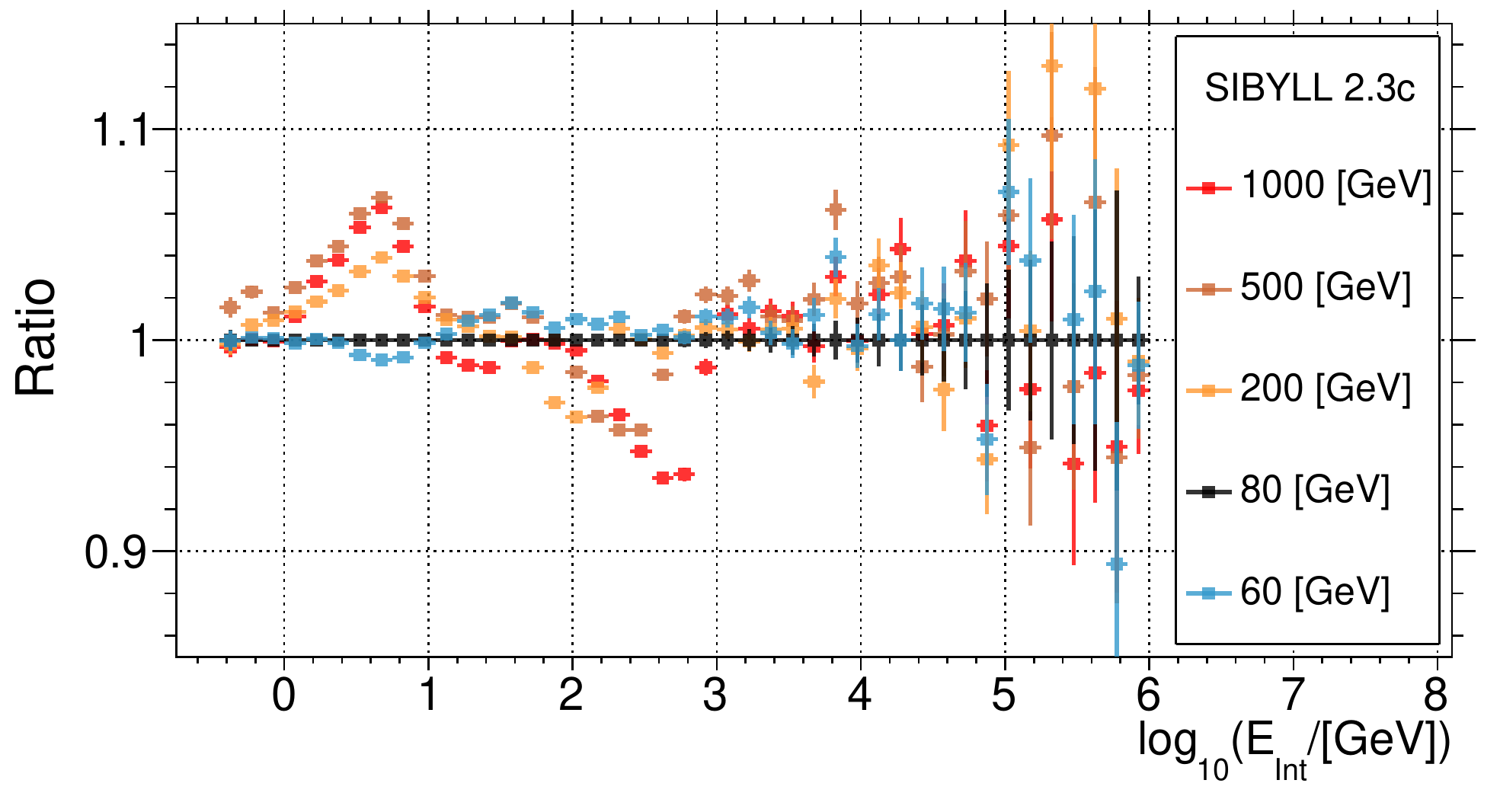}
    \caption{Ratios between hadronic interaction distributions in PeV proton initiated events with different transition energy values with respect to the default of 80\,GeV.}
    \label{EQSSwRelDiffthemselves}
\end{figure}

The low energy model and the choice of the transition energy  have been shown to significantly influence the ground level outcome in simulated events (eg. Figures \ref{LP_proton_pion}, \ref{GLmu_vs_KLPprotons} and \ref{MM_muons_pt_K}). In this section we investigate the effect of strengthening or weakening the low energy model role in PeV proton initiated events by comparing hadronic interaction rates in showers generated with diverse transition energy values.   

In Figure \ref{DiffswEHEmodels} we showed how the default transition energy distributions differed between EPOS-LHC and SIBYLL 2.3c and QGSJetII-04. In Figure \ref{EQSSwRelDiffthemselves}, the ratio between the number of hadronic interactions in the shower for different transition energies is shown.  By increasing the transition energy and therefore the role of UrQMD in the shower, the distributions tend towards agreement by a decrease of EPOS-LHC's and increase of SIBYLL 2.3c's and QGSJetII-04's number. The variations are found to  be  up to 10\%  when the low energy model rules approximately 97\% of the shower interactions (transition energy at 1\,TeV). This opposite behaviour between models reveals the UrQMD distribution lying in between EPOS-LHC's and SIBYLL's. Nevertheless, recall that if the agreement is sought by increasing the transition energy, the high energy model phenomenological definition of the shower will be significantly suppressed.

Lastly an important remark derived from the behaviour observed in Figure \ref{EQSSwRelDiffthemselves}. The only high energy model character that can be enhanced by a modification of the transition energy is EPOS-LHC's. When significantly decreasing the transition energy ($> 60$\,GeV) and giving more importance to the dynamics of the high energy model, QGSJetII-04 lowers the number of collisions resulting in a slight loss of interactions by which the model defines the shower. SIBYLL 2.3c fails to generate events at such low energies.  In contrast, EPOS-LHC presents a slight increase in the number of interactions when the transition energy is lowered, entailing a larger influence of the high energy model physics. 
\end{document}